\documentclass[traditabstract]{aa} 
\usepackage{graphicx}
\usepackage{txfonts}
\usepackage{natbib}
\usepackage{hyperref}
\begin{document}
   \title{A study of the metal-rich, thick disk globular cluster NGC\,5927 {\thanks{Based on observations collected at
           the European Southern Observatory, Chile, ESO, Programme ID 079.B-0721(A)}}}

   \subtitle{A stellar inventory}

   \author{J. Simmerer
          \inst{1}\inst{2}
          \and
          S. Feltzing\inst{1}
          \and
          F. Primas\inst{3}
          }

   \institute{
	    Lund Observatory, Box 43, SE-221 00 Lund, Sweden\\
              \email{sofia@astro.lu.se} 
              \and
University of Utah Department of Physics and Astronomy, Salt Lake City, UT 84112 \\
              \email{jennifer@physics.utah.edu}
	 \and 
	     European Southern Observatory, Karl-Schwarzschild Str. 2, 85748 Garching b. M\"unchen, Germany \\
	     \email{fprimas@eso.org}
             }

   \date{Received 2012-11-14; accepted 2013-04-23}

 
   \abstract {We present a list of 72 radial velocity member stars in
     the metal-rich globular cluster NGC\,5927. The radial velocities
     are based on multi-epoch, multi-fibre spectra. We identify 46
     RGB/HB stars and 26 turn-off stars that are radial velocity
     members in the cluster. This cluster is situated quite close to
     the disk and hence fore- and/or background contamination,
     especially in the outskirts of the cluster, can be quite
     severe. Fortunately, the cluster has a radial velocity (we
     determine it to v$_r$=--104.03$\pm$5.03 km s$^{-1}$) that sets it
     clearly apart from the bulk velocities of the surrounding,
     background, and foreground stellar populations. Hence, our
     identification of members is clean and we can quantify a 50\%
     contamination when stars in the outer part of the cluster are
     selected solely based on position in the colour-magnitude diagram
     as opposed to selections based on radial velocities. }


   \maketitle
%
\section{Introduction}

It has long been recognized that the globular cluster system in the
Milky Way contains at least two distinct groups: a metal-poor system
and a metal-rich system \citep{zinn1985}. While some individual
clusters might be accreted from satellite systems, metal-poor clusters
are typically associated with the halo \citep{bica2006, Law2010}.  The
metal-rich sub-system is strongly concentrated around and therefore
typically associated with the Galactic bulge \citep{bica2006}.
Additional structure in the metal rich-system was originally proposed
by \citet{armandroff1989}, who suggested that these clusters were
kinematically associated with the stellar thick disk. Proper motion
studies confirm Bulge membership for some metal-rich clusters for
which orbits can be derived \citep{Dinescu1999b, Dinescu1999a,
  dinescu2003, Casetti-Dinescu2007, Casetti-Dinescu2010}.  However, a
few clusters may still be associated with the thick disk.  One of the
best thick disk candidate clusters is \object{NGC\,5927}
\citep{Casetti-Dinescu2007}.

No detailed spectroscopic abundance analysis of NGC\,5927 has been
published in the last 20 years despite evidence that the overall
metallicity of the cluster is poorly constrained.  Based on Ca\,{\sc ii}
IR triplet measurements by \citet{rutledge1997}, three different
calibrations yield varying results: --0.34 dex (Zinn-scale), --0.64
dex (Carretta \& Gratton -scale), and --0.67 dex
\citep{kraft2003}. \citet{2003MNRAS.339..897T} used Lick indices to
derive [Fe/H] = --0.21$\pm$0.11.  

At $l\simeq 34^{\circ}$ contamination from the background (Bulge)
population is of small concern and the foreground contamination is
expected to be moderate (but as we shall see it is still significant).
This property has made NGC\,5927 a popular empirical fiducial cluster
in studies of extra-galactic resolved stellar populations and it is
often used as a template for colour-magnitude diagrams, integrated
colours, and spectra of metal-rich, single age stellar
populations. Recent examples include \citet{brown2004,brown2005}
\citet{Bedin2005}, and \citet{2004A&A...428...79S}.

For these reasons it is interesting to provide a clean sample of
members of the cluster that can be used for further detailed studies,
e.g., as targets for high-resolution spectroscopic studies. In this
paper we present a VLT-FLAMES observation of a large number of stars
in this cluster and derive the radial velocities of the stars as well
as the systemic velocity of the cluster.

%
\section{Selection of targets}

\begin{figure*}
\centering
\includegraphics[width=13cm]{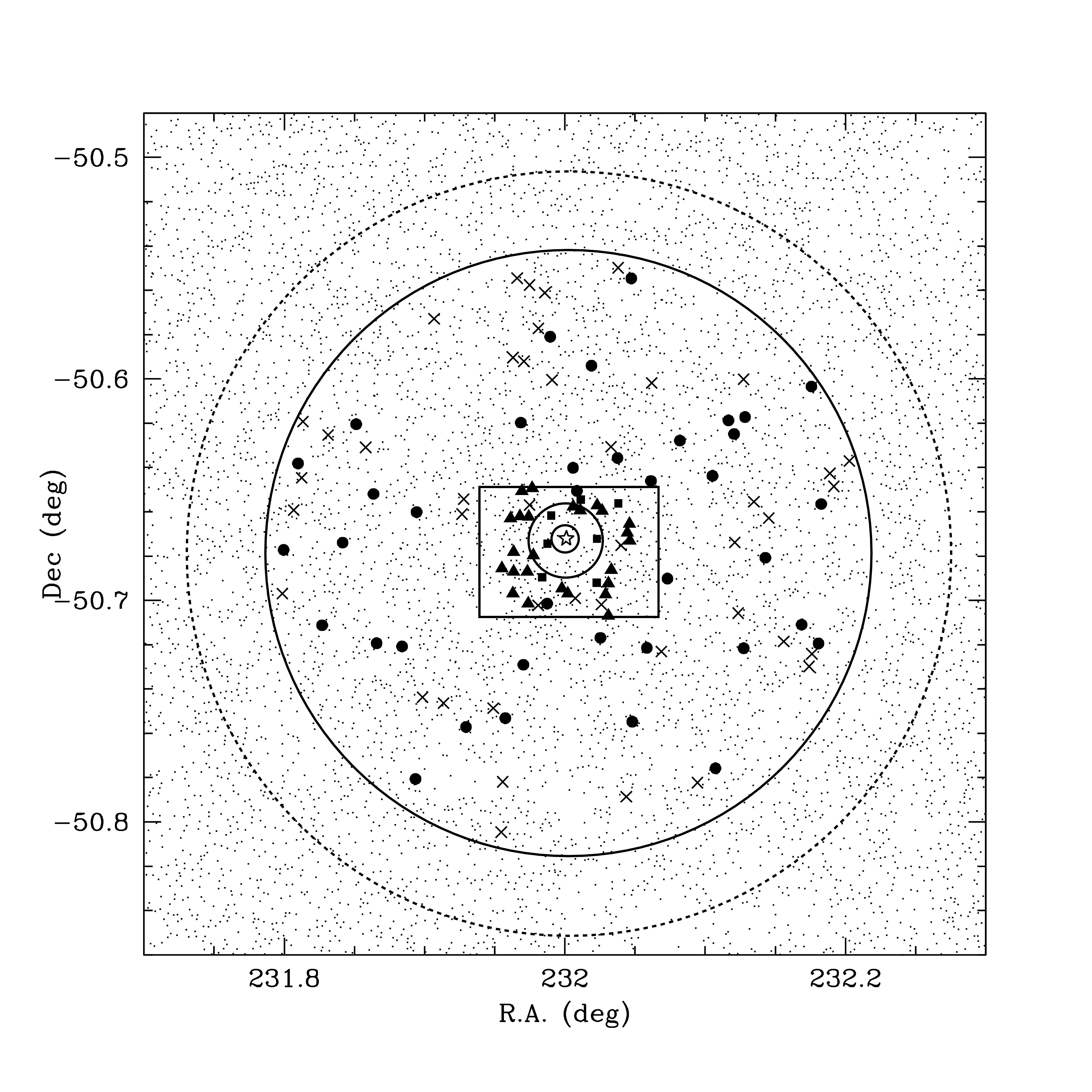}
\caption{Position on the sky of all the fibres. Filled squares mark
  HB/RGB-UVES stars with radial velocity membership.  RGB-GIRAFFE as
  and TO radial velocity members are marked with filled circles and
  filled triangles, respectively. Crosses mark all stars that are
  non-members according to their radial velocities. The core and
  half-mass radii are marked as circles close to the centre of the
  cluster (which is marked by an open star). The tidal radius is also
  marked with a circle with a dashed line. Data for the various radii
  were taken from the latest web-based version of \cite{harris1996}.
  The FLAMES field of view is indicated by a large solid circle. The
  square shows the rough position of the ACS camera observations.}
\label{fig:pos}
\end{figure*}

\begin{figure*}
\centering
\includegraphics[width=13cm]{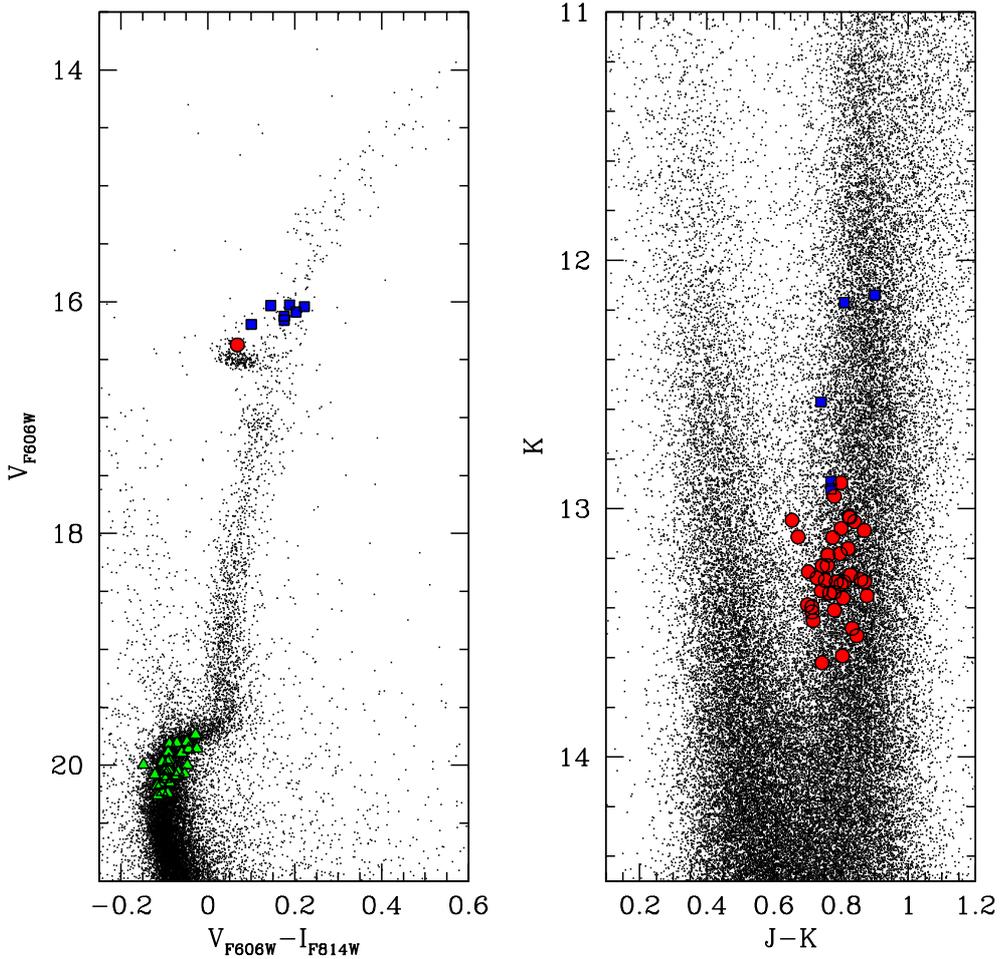}
\caption{Position of program stars in color-magnitude diagrams. {\bf
    Left:} RGB-UVES (light blue squares), RGB-GIRAFFE (red circles),
  and TO stars (green triangles) from HST ACS photometry. {\bf Right:}
  RGB-UVES and RGB-GIRAFFE stars from 2MASS photometry. Black dots indicate
the full CMDs.}
\label{cmdfig}
\end{figure*}

\begin{table*}
\caption{Basic data for candidate HB and RGB stars observed with UVES ($R\sim 40\,000$) in NGC 5927.\label{hb.tab}}  
\setlength{\tabcolsep}{0.02in}
\begin{tabular}{llrrrrrrrrrrr} 
\hline\hline             
\noalign{\smallskip}
Star        & 2MASS ID  & R.A.          & Dec.           &  v$_{Helio}$  & $\sigma$ & $V_{\rm 606}$   &  $I_{\rm 814}$ &  $J$  &  $H$    &  $K_S$   \\
   & & (fibre)       & (fibre)        &  [km\,s$^{-1}$] & [km\,s$^{-1}$]  & \\
\noalign{\smallskip}
\hline                        
\noalign{\smallskip}
\object{NGC 5927-01} & 15275603-5041223 & 15 27 56.16 & --50 41 22.20  & --102.01 & 0.37 & 16.02 & 15.83 &   13.355 & 12.619 & 12.565  \\
\object{NGC 5927-02} &                  & 15 27 56.88 & --50 40 28.20  & --94.39  & 0.19 & 16.12 & 15.94 &     &  & \\
\object{NGC 5927-03} & 15275764-5039423 & 15 27 57.60 & --50 39 42.12  & --101.05 & 0.36 & 16.16 & 15.98 &   13.707 & 13.021 & 12.891  \\
\object{NGC 5927-04} & 15280273-5039163 & 15 28 02.64 & --50 39 16.20  & --105.61 & 0.52 & 16.03 & 15.88 &    & 12.282 & 12.129  \\
\object{NGC 5927-05} & 15280546-5040214 & 15 28 05.52 & --50 40 21.72  & --105.37 & 0.46 & 16.04 & 15.82 &   13.026 & 12.432 & 12.161  \\
\object{NGC 5927-06} & 15280537-5041317 & 15 28 05.28 & --50 41 31.56  & --102.34 & 0.44 & 16.19 & 16.09 &   13.738 & 13.012 & 12.917  \\
\object{NGC 5927-07} &                  & 15 28 09.12 & --50 39 22.32  & --105.84 & 0.56 & 16.08 & 15.88 &    &  &  \\
\noalign{\smallskip}
\hline                        
\end{tabular}\newline
\footnotesize{Column one the IDs used by us for the stars while column two gives the 2MASS 
  identification. 
  Columns three and four contain the target fibre coordinates, column five the
  heliocentric radial velocity,  and column six
  the radial velocity standard deviation as derived in Sect.\,\ref{Sect:rv}. Columns seven and eight
  list the in-flight ACS magnitudes (T. Brown {\it priv. com.}).
  The last three columns list the
  $JHK_S$ magnitudes from 2MASS.}  
\end{table*}

   \begin{table*}
\caption[]{Basic data for candidate TO stars observed with GIRAFFE in NGC 5927.}
\label{to.tab}
\begin{tabular}{lllllllllllllllll}
\hline \hline
\noalign{\smallskip}
 Star   & R.A.        & Dec.          & R.V.    &  $V_{\rm 606}$  & $(V_{\rm 606}-I_{\rm 814})$   \\
        & (fibre)     & (fibre)       & [km\,s$^{-1}$]  &   \\
\noalign{\smallskip}
\hline
\noalign{\smallskip}          
 TO001 &15 27 49.25 & --50 41 08.00  & --103.74  & 20.11 &     --0.097      \\
 TO004 &15 27 50.70 & --50 39 46.40 & --107.32  & 20.23 &      --0.105    \\
 TO012 &15 27 51.21 & --50 40 41.10 & --106.90  & 19.81 &      --0.088  \\
 TO013 &15 27 52.64 & --50 39 02.20  & --98.06   & 20.05 &     --0.101     \\
 TO014 &15 27 51.24 & --50 41 13.50 & --99.88   & 20.24 &      --0.092     \\
 TO015 &15 27 52.31 & --50 39 43.10 & --96.88   & 19.80 &      --0.048     \\
 TO016 &15 27 51.12 & --50 41 48.70 & --118.28  & 20.16 &      --0.109     \\
 TO021 &15 27 53.81 & --50 39 44.20 & --111.05  & 19.97 &      --0.104     \\
 TO022 &15 27 54.42 & --50 38 57.60 & --102.84  & 19.89 &      --0.060     \\
 TO033 &15 27 53.58 & --50 41 13.30 & --110.28  & 20.04 &      --0.097     \\
 TO036 &15 27 54.61 & --50 40 46.50 & --105.18  & 19.99 &      --0.148     \\
 TO037 &15 27 53.72 & --50 42 05.10  & --99.15   & 19.85 &     --0.025     \\
 TO051 $\dagger$ &15 27 59.47 & --50 41 40.70 & --105.33  &        &                  \\
 TO052 &15 28 1.45  & --50 39 27.30 & --102.29  & 20.04 &      --0.090     \\
 TO054 &15 28 0.52  & --50 41 49.00 & --105.52  & 20.05 &      --0.068     \\
 TO056 &15 28 2.62  & --50 39 33.50 & --110.10  & 20.09 &      --0.101     \\
 TO064 &15 28 5.51  & --50 39 25.70 & --99.61   & 19.88 &      --0.091     \\
 TO070 &15 28 6.40  & --50 39 34.60 & --104.23  & 20.00 &      --0.047     \\
 TO083 &15 28 6.97  & --50 41 50.20 & --106.09  & 20.15 &      --0.088     \\
 TO084 &15 28 7.43  & --50 41 32.50 & --105.61  & 20.07 &      --0.053     \\
 TO086 &15 28 7.93  & --50 41 10.50 & --103.21  & 19.97 &      --0.092     \\
 TO088 &15 28 7.43  & --50 42 24.90 & --93.08   & 20.14 &      --0.085     \\
 TO090 &15 28 9.62  & --50 40 30.50 & --103.40  & 20.26 &      --0.115     \\
 TO101 &15 28 10.67 & --50 40 10.00 & --107.09  & 19.80 &      --0.071     \\
 TO103 &15 28 11.03 & --50 39 55.70 & --108.89  & 20.08 &      --0.122     \\
 TO106 &15 28 11.14 & --50 40 22.60 & --95.77   & 19.95 &      --0.089     \\
\noalign{\smallskip}
\hline
\noalign{\smallskip}          
 TO020    & 15 27 53.96 & --50 39 24.90 &       13.82  &  19.74  &  --0.028            \\
 TO042    & 15 27 55.43 & --50 42  8.20 &       15.37  &  20.09  &  --0.073      \\
 TO058    & 15 28  1.78 & --50 41 56.60 &        9.71  &  19.86  &  --0.044      \\
 TO078    & 15 28  6.25 & --50 42  6.80 &       32.98  &  20.18  &  --0.117       \\
            \hline
         \end{tabular}\newline
\footnotesize{Fibre coordinates are given in columns two and three.  Column four
contains the heliocentric radial velocity derived from the combined spectrum (see Sect. \ref{Sect:rv}).
The last columns contain the in-flight ACS magnitudes (T. Brown {\it priv. com.}).
The first 26 lines contain the stars with a
  radial velocity that makes them likely members of NGC 5927. The last four
  entries are the stars that fall more than 3\,$\sigma$ outside the
  mean cluster radial velocity.\newline
 $\dagger$ Star T051 has no detection in the Brown photometry within 1.5 arcsec.  There are
two sources within 1.7 arcseconds. We cannot unambiguously associate colors from
\citet{brown2005} with our target source.} 
\end{table*}

\begin{table*}
\caption{Basic data for candidate RGB stars observed with GIRAFFE in NGC 5927.}  
\setlength{\tabcolsep}{0.02in}
\begin{tabular}{ccllccc|ccllccc} 
\hline\hline             
\noalign{\smallskip}
\multicolumn{7}{c}{Cluster members} &\multicolumn{7}{c}{Cluster non-members}\\
Star  & 2MASS ID           & R.A.          & Dec            & $J$     & $H$    &  $K_S$& Star  & 2MASS ID           & R.A.          & Dec            & $J$     & $H$    &  $K_S$ \\
\noalign{\smallskip}
\hline                        
\noalign{\smallskip}
M003  &  15280206-5039020  &  15 28  2.06  &  -50 39  2.00  &  14.08  &  13.39  &  13.3   &M130  &  15274240-5039396  &  15 27 42.40  &  -50 39 39.60  &  14.21  &  13.61  &  13.38  \\
M012  &  15275697-5042051  &  15 27 56.97  &  -50 42  5.10  &  13.88  &  13.17  &  13.08  &M138  &  15274271-5039153  &  15 27 42.71  &  -50 39 15.30  &  13.81  &  13.06  &  12.98  \\
M061  &  15280140-5038244  &  15 28  1.40  &  -50 38 24.40  &  14.1   &  13.43  &  13.27  &M171  &  15281652-5043230  &  15 28 16.52  &  -50 43 23.00  &  14.25  &  13.52  &  13.38  \\
M098  &  15280608-5043008  &  15 28  6.08  &  -50 43  0.80  &  14.17  &  13.56  &  13.36  &M211  &  15282906-5040257  &  15 28 29.06  &  -50 40 25.70  &  13.99  &  13.31  &  13.16  \\
M111  &  15280900-5038085  &  15 28  9.00  &  -50 38  8.50  &  14.36  &  13.6   &  13.51  &M213  &  15275781-5036016  &  15 27 57.81  &  -50 36  1.60  &  13.6   &  12.86  &  12.72  \\
M115  &  15281754-5041246  &  15 28 17.54  &  -50 41 24.60  &  14.23  &  13.53  &  13.35  &M222  &  15274775-5044550  &  15 27 47.74  &  -50 44 55.00  &  14.4   &  13.7   &  13.53  \\
M120  &  15281471-5038457  &  15 28 14.71  &  -50 38 45.70  &  13.95  &  13.27  &  13.19  &M227  &  15282965-5042205  &  15 28 29.65  &  -50 42 20.50  &  14.06  &  13.48  &  13.24  \\
M126  &  15280789-5037501  &  15 28  7.89  &  -50 37 50.10  &  13.97  &  13.33  &  13.19  &M232  &  15281484-5036067  &  15 28 14.84  &  -50 36  6.70  &  14.27  &  13.62  &  13.49  \\
M151  &  15275290-5043446  &  15 27 52.90  &  -50 43 44.60  &  14.16  &  13.4   &  13.29  &M238  &  15283233-5039196  &  15 28 32.33  &  -50 39 19.60  &  14.1   &  13.53  &  13.44  \\
M152  &  15281401-5043171  &  15 28 14.02  &  -50 43 17.10  &  14.1   &  13.47  &  13.3	  &M240  &  15275299-5035315  &  15 27 52.99  &  -50 35 31.50  &  13.67  &  12.96  &  12.9   \\
M160  &  15275246-5037111  &  15 27 52.46  &  -50 37 11.10  &  14.09  &  13.49  &  13.39  &M250  &  15275112-5035252  &  15 27 51.13  &  -50 35 25.20  &  13.84  &  13.21  &  13.14  \\
M193  &  15281966-5037402  &  15 28 19.67  &  -50 37 40.20  &  13.73  &  13.15  &  12.95  &M253  &  15273924-5044470  &  15 27 39.24  &  -50 44 47.00  &  14.01  &  13.41  &  13.33  \\
M199  &  15273464-5039364  &  15 27 34.64  &  -50 39 36.40  &  13.96  &  13.36  &  13.25  &M256  &  15283486-5039461  &  15 28 34.86  &  -50 39 46.10  &  14.11  &  13.4   &  13.26  \\
M204  &  15282525-5038375  &  15 28 25.25  &  -50 38 37.50  &  14.13  &  13.46  &  13.42  &M264  &  15273567-5044373  &  15 27 35.67  &  -50 44 37.30  &  13.72  &  12.97  &  12.83  \\
M228  &  15274983-5045114  &  15 27 49.82  &  -50 45 11.40  &  14.19  &  13.5   &  13.41  &M276  &  15275546-5034378  &  15 27 55.47  &  -50 34 37.80  &  14.39  &  13.64  &  13.65  \\
M230  &  15280455-5035385  &  15 28  4.55  &  -50 35 38.50  &  14.32  &  13.54  &  13.48  &M285  &  15272593-5037513  &  15 27 25.93  &  -50 37 51.30  &  13.99  &  13.33  &  13.22  \\
M235  &  15281153-5045171  &  15 28 11.53  &  -50 45 17.10  &  13.86  &  13.22  &  13.04  &M291  &  15283744-5043066  &  15 28 37.44  &  -50 43  6.60  &  13.83  &  13.2   &  13.14  \\
M244  &  15273214-5043148  &  15 27 32.14  &  -50 43 14.80  &  14.17  &  13.51  &  13.45  &M295  &  15283057-5036006  &  15 28 30.57  &  -50 36  0.60  &  13.68  &  12.99  &  12.87  \\
M248  &  15283429-5040506  &  15 28 34.29  &  -50 40 50.60  &  14.14  &  13.45  &  13.29  &M297  &  15274935-5046549  &  15 27 49.35  &  -50 46 54.90  &  13.84  &  13.19  &  13.01  \\
M252  &  15282894-5037295  &  15 28 28.94  &  -50 37 29.50  &  13.86  &  13.21  &  13.03  &M303  &  15275659-5033401  &  15 27 56.58  &  -50 33 40.10  &  13.78  &  13.18  &  13.05  \\
M255  &  15283059-5043177  &  15 28 30.59  &  -50 43 17.70  &  14.36  &  13.64  &  13.62  &M304  &  15281049-5047189  &  15 28 10.50  &  -50 47 18.90  &  13.99  &  13.25  &  13.11  \\
M258  &  15272726-5039069  &  15 27 27.26  &  -50 39  6.90  &  13.89  &  13.29  &  13.12  &M320  &  15275396-5033273  &  15 27 53.96  &  -50 33 27.30  &  14.4   &  13.72  &  13.53  \\
M260  &  15282799-5037075  &  15 28 27.99  &  -50 37  7.50  &  14.11  &  13.47  &  13.4	  &M321  &  15284225-5043270  &  15 28 42.25  &  -50 43 27.00  &  14.17  &  13.54  &  13.48  \\
M263  &  15274309-5045256  &  15 27 43.09  &  -50 45 25.60  &  13.98  &  13.3   &  13.16  &M322  &  15273763-5034221  &  15 27 37.63  &  -50 34 22.10  &  13.96  &  13.28  &  13.1   \\
M267  &  15275749-5034515  &  15 27 57.49  &  -50 34 51.50  &  13.99  &  13.36  &  13.23  &M323  &  15271949-5037310  &  15 27 19.49  &  -50 37 31.10  &  13.89  &  13.18  &  13.07  \\
M270  &  15272780-5043096  &  15 27 27.80  &  -50 43  9.60  &  14.11  &  13.44  &  13.34  &M327  &  15284176-5043473  &  15 28 41.76  &  -50 43 47.30  &  14.36  &  13.76  &  13.7   \\
M273  &  15283082-5037019  &  15 28 30.82  &  -50 37  1.90  &  14.04  &  13.44  &  13.29  &M328  &  15282266-5046563  &  15 28 22.66  &  -50 46 56.30  &  14.22  &  13.5   &  13.38  \\
M284  &  15272199-5040260  &  15 27 21.99  &  -50 40 26.00  &  14.01  &  13.36  &  13.28  &M330  &  15284533-5038336  &  15 28 45.34  &  -50 38 33.60  &  13.87  &  13.23  &  13.02  \\
M299  &  15284050-5042394  &  15 28 40.50  &  -50 42 39.40  &  14.12  &  13.45  &  13.34  &M333  &  15284603-5038547  &  15 28 46.03  &  -50 38 54.70  &  14.16  &  13.51  &  13.38  \\
M300  &  15272430-5037136  &  15 27 24.30  &  -50 37 13.60  &  13.79  &  13.22  &  13.11  &M335  &  15275184-5033162  &  15 27 51.84  &  -50 33 16.20  &  13.89  &  13.28  &  13.19  \\
M305  &  15284385-5039233  &  15 28 43.85  &  -50 39 23.30  &  13.96  &  13.24  &  13.09  &M341  &  15271362-5039331  &  15 27 13.62  &  -50 39 33.10  &  14.22  &  13.57  &  13.49  \\
M308  &  15271847-5042403  &  15 27 18.48  &  -50 42 40.30  &  13.98  &  13.33  &  13.18  &M349  &  15280909-5032594  &  15 28  9.09  &  -50 32 59.40  &  14.36  &  13.78  &  13.54  \\
M319  &  15282577-5046327  &  15 28 25.77  &  -50 46 32.70  &  14.11  &  13.42  &  13.3   &M356  &  15271166-5041490  &  15 27 11.66  &  -50 41 49.10  &  13.73  &  12.98  &  12.85  \\
M326  &  15284338-5043099  &  15 28 43.38  &  -50 43  9.90  &  13.7   &  13.13  &  13.05  &M360  &  15274912-5048166  &  15 27 49.12  &  -50 48 16.60  &  14.25  &  13.56  &  13.4   \\
M339  &  15281136-5033166  &  15 28 11.36  &  -50 33 16.60  &  13.97  &  13.38  &  13.23  &M362  &  15284867-5038134  &  15 28 48.67  &  -50 38 13.40  &  14.22  &  13.54  &  13.38  \\
M342  &  15273444-5046503  &  15 27 34.44  &  -50 46 50.30  &  14.4   &  13.74  &  13.59  &M383  &  15271519-5037093  &  15 27 15.19  &  -50 37  9.30  &  13.82  &  13.15  &  13.02  \\
M351  &  15271434-5038173  &  15 27 14.34  &  -50 38 17.30  &  13.89  &  13.26  &  13.05  &      &                    &               &                &         &         &         \\
M354  &  15271189-5040378  &  15 27 11.89  &  -50 40 37.80  &  13.7   &  13.02  &  12.9   &      &                    &               &                &         &         &         \\
M357  &  15284217-5036124  &  15 28 42.17  &  -50 36 12.40  &  14.07  &  13.41  &  13.33  &      &                    &               &                &         &         &         \\
\noalign{\smallskip}
\hline                        
\end{tabular}\newline
\footnotesize{Columns two and nine are the 2MASS identifications for target RGB stars.  
Columns three, four, ten and eleven
  give the R.A. and Dec. of the GIRAFFE fibre used to observe the star.  
The final columns are the 2MASS $JHK_S$ 
  magnitudes.} 
\label{rgb.tab}     
\end{table*}

The selection of targets in NGC\,5927 was done for three different
types of stars: horizontal branch (HB) stars, red giant branch (RGB)
stars, and turn-off (TO) stars. Seven bright HB and RGB stars
were observed at high resolution (R$\sim 40\,000$) and will be
referred to as HB/RGB-UVES.  The remaining RGB targets were observed
at lower resolution (R$\sim 20\,000$) and will be referred to as
RGB-GIRAFFE.  Target selection was based on photometric color cuts and
designed to maximize the number of fibres allocated to stars.

\paragraph{HB and RGB stars for high resolution fibres (HB/RGB-UVES)}
were selected using a colour-magnitude diagram (CMD) based on
observations with the Advanced Camera for Surveys (ACS) on-board
Hubble. The selection of HB/RGB-UVES stars was straightforward. The
ACS CMD allowed us to select targets in the crowded core of the
cluster, increasing our chances of observing cluster members (compare
Fig.\,\ref{cmdfig}). \footnote{At the time of selection this
  photometry was unpublished. Later a refereed publication of this
  data was published in \citep{brown2005}. T. Brown has kindly
  provided us with both the calibrated ACS photometry and his
  astrometric solution. Our target cross-match was done in RA and Dec.
  We found target matches for every star on the HB/RGB}.

The final list of HB/RGB-UVES targets is given in Table\,\ref{hb.tab}.
These targets are bright enough to have 2MASS NIR photometry
\citep{skrutskie2006}. $JHK_S$ magnitudes are included in the table if
their respective quality flag is at least ``B''. NGC\,5927-02 and -07
have no reliable 2MASS photometry and NGC\,5927-05 and NGC\,5927-06
have contamination flagged for its $JHK_S$ magnitudes.

\paragraph{TO stars} were selected from the same ACS CMD as the
HB/RGB-UVES stars. A small box was placed around the TO in the CMD.
In total 30 suitable TO star were selected. These are listed in
Table\,\ref{to.tab} and their positions in the CMD are shown in
Fig.\,\ref{fig:pos}. \footnote{We also cross-matched these
stars with the final photometry and astrometry from T.Brown 
and found that all stars apart from T051 have a unique match.}

\paragraph{RGB stars for low-resolution fibres (RGB-GIRAFFE)} were
selected from a CMD based on 2MASS photometry
\citep{skrutskie2006}. These targets were selected to be further from
the core of the cluster in order facilitate the positioning of the
fibres, but also to probe the cluster properties to larger radii.  The
stars were selected to roughly match the magnitudes of the HB/RGB-UVES
stars with $13.6< K_{S} <12.8$ (compare CMD in Fig.\,\ref{cmdfig}).
In total 75 RGB-GIRAFFE candidate stars were selected and observed.
They are listed in Table \ref{rgb.tab} along with the associated fibre
coordinate, the 2MASS identification, and the $JHK_S$ magnitudes.

The position of all targets on the sky is shown in
Fig.\,\ref{fig:pos}.  Since our targets are drawn from two different
photometric sources, our fibre coordinate system represents a
compromise between two astrometric systems.  For the sake of
presenting a single internally consistent coordinate system and aiding
future cross-identification, we list the R.A. and Dec. for the fibre
positions in Tables\,\ref{hb.tab},\ref{to.tab}, and \ref{rgb.tab}
rather than the astrometry used in the selection.  Fibre coordinates
are within 0.5$''$ of the ACS coordinates \citep{brown2005} and within
0.13$''$ (RGB-GIRAFFE targets) or 1.0$''$ (HB/RGB-UVES targets) of the
2MASS coordinates.

%
\section{Observations and data reduction}
\label{Sect:obs}

\begin{table*}
  \caption{Observing dates, exposure times, average seeing, airmass, and helio-centric corrections
    for all observations. Note that the OBs are listed in the order they were 
    executed. \label{obs.tab}     
  }            
\begin{tabular}{llllllllll} 
\hline\hline             
\noalign{\smallskip}
Observing & Date & UT & JD & Exp-time & Average& Airmass & \multicolumn{1}{c}{Helio-centric} \\ 
Block  &   && & [s]& seeing & &\multicolumn{1}{c}{correction}\\
& & && & during obs. & &\multicolumn{1}{c}{[km\,s$^{-1}$]}\\
& & && & [''] & &\\
\noalign{\smallskip}
\hline                        
\noalign{\smallskip}
281555 & 2007 05 12 & 02:48:23.543  & 2454232.62184  & 3600  &1.16$\pm$0.45 &     1.22 &   5.74 \\
281556 & 2007 05 12 & 05:43:13.406  & 2454232.74325  & 3600  &1.27$\pm$0.60 &     1.13 &   5.49 \\
281557 & 2007 05 12 & 06:50:11.231  & 2454232.78976  & 3600  &1.23$\pm$0.54 &     1.22 &   5.39 \\
281558 & 2007 05 12 & 07:52:11.744  & 2454232.83282  & 3600  &1.24$\pm$0.58 &     1.37 &   5.32 \\
281559 & 2007 05 13 & 01:32:58.649  & 2454233.56949  & 3600  &0.92$\pm$0.01 &     1.42 &   5.40 \\
281560 & 2007 05 15 & 00:05:47.728  & 2454235.50897  & 3600  &0.59$\pm$0.01 &     1.88 &   4.62 \\
281561 & 2007 05 15 & 01:11:22.661  & 2454235.55452  & 3600  &0.42$\pm$0.01 &     1.48 &   4.56 \\
281562 & 2007 05 17 & 02:03:43.510  & 2454237.59090  & 3600  &0.86$\pm$0.01 &     1.28 &   3.64 \\
281563 & 2007 05 20 & 00:49:43.705  & 2454240.53954  & 3600  &1.15$\pm$0.44 &     1.49 &   2.42 \\
281564 & 2007 05 22 & 02:53:50.236  & 2454242.62574  & 3600  &1.47$\pm$0.87 &     1.16 &   1.40 \\
281565 & 2007 05 22 & 03:58:41.221  & 2454242.67077  & 3600  &1.54$\pm$0.93 &     1.11 &   1.30 \\
281566 & 2007 06 17 & 05:09:48.056  & 2454268.71979  & 3600  &1.24$\pm$0.57 &     1.31 &  --9.77 \\
281567 & 2007 06 22 & 00:38:31.127  & 2454273.53123  & 3600  &2.13$\pm$1.62 &     1.17 & --11.36 \\
281568 & 2007 06 22 & 01:45:56.216  & 2454273.57805  & 3600  &1.58$\pm$1.00 &     1.12 & --11.45 \\
281569 & 2007 06 22 & 02:52:30.930  & 2454273.62428  & 3600  &1.26$\pm$0.59 &     1.13 & --11.55 \\
281570 & 2007 06 22 & 04:04:06.944  & 2454273.67400  & 3600  &1.24$\pm$0.56 &     1.21 & --11.64 \\
281574 & 2007 07 19 & 00:37:24.431  & 2454300.52901  & 3600  &0.72$\pm$0.01 &     1.11 & --20.20 \\
281573 & 2007 07 20 & 00:40:29.476  & 2454301.53108  & 3600  &1.36$\pm$0.74 &     1.12 & --20.46 \\
281572 & 2007 07 20 & 01:42:01.361  & 2454301.57381  & 3600  &0.89$\pm$0.01 &     1.16 & --20.54 \\
281571 & 2007 07 22 & 00:48:44.083  & 2454303.53667  & 3600  &1.24$\pm$0.56 &     1.12 & --20.97 \\
\noalign{\smallskip}
\hline                        
\end{tabular}\newline
\footnotesize{Column one lists the name of the observing block, column
  two the date the observations were taken, columns three and
  four give the starting time of the observation (in UT) 
  and the Julian date, column five lists the exposure time, column six the 
  average seeing during the observation including a $\sigma$ 
  and
  column seven the airmass at which the observations were carried out. The last column lists the helio-centric
  corrections derived in this work and applied when combining spectra form each of the 20 OBs into
  one single spectrum.}
\end{table*}

\begin{figure}
\centering
\includegraphics[width=8cm]{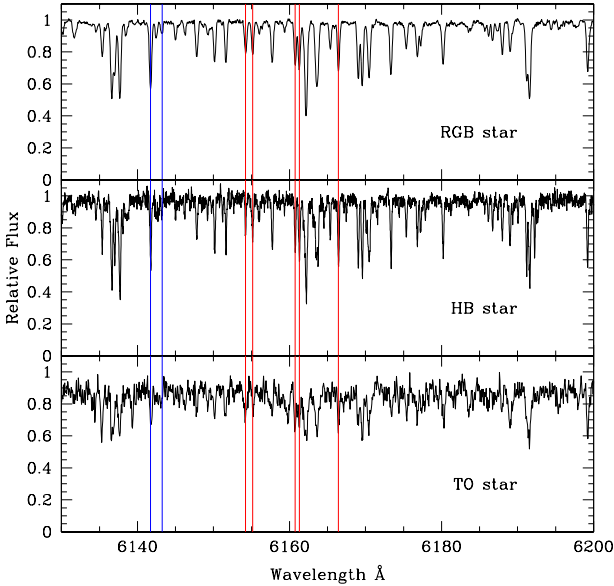}
\caption{A section of three of our target spectra: M151, NGC 5927-05,
  and T021. The spectra are arranged in order of increasing S/N.
  Two blue lines (to the left) indicate the location of the Ba\,{\sc ii} and
  Zr\,{\sc i} lines at 614.17 nm and 614.31 nm, respectively.  The five
   red lines indicate the position of the Si\,{\sc i} line at
  615.51\,nm, the Na\,{\sc i} lines at 615.42\,nm and 616.07\,nm, and the
  Ca\,{\sc i} lines at 616.13\,nm, and 616.64\,nm, respectively.  }
\label{fig:spectra}
\end{figure}

For the seven high-resolution (R$\sim$40,000) fibres allocated to
HB/RGB-UVES targets we used the setting centered at 580\,nm which
covers the spectrum between 480.0\,nm and 682.0\,nm with a gap between
577.5 and 582.5\,nm (CCD\#3, filter SHP700, red arm only).  For the
105 (R$\sim$20,000) Medusa fibres, we selected the grating setting
HR13. HR13 (612.0 -- 640.5\,nm) includes several astrophysically
interesting elements (e.g. O, Na, and Ba) and still provides adequate
throughput for the fainter TO stars.

Due to the crowding in the central part of the cluster, good
  seeing was imperative and therefore the observations were carried
  out in service mode using the FLAMES spectrograph
  \citep{pasquini2002} mounted on UT2 on Paranal. The seeing
  requirement was set to 1.2$\,{''}$ for the service mode observations
  to minimize the loss of light due to finite fibre size and at the
  same time maximize the amount of useful observing time
  available. The explicit aim was, for each observing block (OB), to
  achieve S/N=11 @630nm for HB/RGB-UVES stars, S/N=4 @620nm for TO
  stars, and S/N=25 for HB-GIRAFFE stars. Each OB was observed for
  3600\,s. In total we observed 20 OBs. Observations were carried out
  in service mode during May and July 2007. The observing conditions
  for each OB are listed in Table\,\ref{obs.tab}. The estimated S/N
were achieved.

Initially, the data were reduced as part of the ESO service mode
operations using the GIRAFFE pipeline version 1.0 \footnote{See
  {\url{http://www.eso.org/sci/software/pipelines/}} for a listing of
  the GIRAFFE pipeline versions.}. Upon inspection it soon became
clear that what is known as the ``CCD glow'' was a serious problem for
our spectra. This glow shows itself in the form of a higher background
in one corner of the CCD. In particular it affects fibres with higher
numbers. It was thus necessary to undertake a re-reduction of the
data. This was done using the $\beta$-version of the GIRAFFE pipeline
version 2.3, released specifically to deal with the CCD glow.

Before combining the spectra from the 20 OBs into a final spectrum
with increased S/N the necessary helio-centric corrections, as
calculated within {\sc iraf}\footnote{IRAF is distributed by National
  Optical Astronomy Observatories, operated by the Association of
  Universities for Research in Astronomy, Inc., under contract with
  the National Science Foundation, USA.} were applied (see
Table\,\ref{obs.tab}).  For a few RGB-GIRAFFE stars some spectra were
rejected (see Table\,\ref{rvrgb.tab} and the discussion in
Sect.\,\ref{Sect:ind}). In some cases there were also one out of the
20 spectra that was not usable (e.g., due to a slight off-set in the
placement of the fibre, resulting in a too low flux). The spectra were
combined using the {\sc scombine} task in IRAF, using a min-max
rejection.  A few example spectra are shown in
Figure\,\ref{fig:spectra}.

%
\section{Determination of radial velocities}
\label{Sect:rv}

\begin{table*}
\caption[]{Radial velocities for RGB-GIRAFFE targets. Both member stars in NGC\,5927 (left column) as well as
non-members (right panel) are listed-.
}
\label{rvrgb.tab}
\renewcommand{\footnoterule}{}  
\begin{tabular}{rrlrr|rrlrr}
\hline \hline
\noalign{\smallskip}
\multicolumn{5}{c}{Cluster members} & \multicolumn{5}{c}{Cluster non-members}\\
Star &   R.V.& $\sigma$ & \multicolumn{2}{c}{Comments}  &Star &   R.V.& $\sigma$ &\multicolumn{2}{c}{Comments}\\
&  [km\,s$^{-1}$]  &  [km\,s$^{-1}$] & & & & [km\,s$^{-1}$]  &  [km\,s$^{-1}$]\\
\noalign{\smallskip}
\hline
\noalign{\smallskip}
 M003 & --94.65 & 0.66 &                         &           & M130 & --4.198 & 3.98 &See Sect\,\ref{Sect:ind} &           \\          
 M012 & --98.08 & 0.64 &                         &           & M138 &  +11.24 & 0.89 &                         &           \\          
 M061 & --101.4 & 0.88 &                         &           & M171 & --29.01 & 0.93 &                         &           \\          
 M098 & --104.9 & 0.99 &                         & 1 rej     & M211 & --82.82 & 0.79 &                         &           \\          
 M111 & --109.0 & 1.39 &                         &           & M213 & --51.61 & 0.72 &                         &           \\          
 M115 & --101.8 & 0.74 &                         &           & M222 & --27.92 & 1.08 &                         &           \\          
 M120 & --115.6 & 0.72 &                         &           & M227 &         &      &See Sect\,\ref{Sect:ind} &13 spectra \\            
 M126 & --99.52 & 1.32 &                         & 5 spectra & M232 &  --30.0 & 0.94 &                         &           \\          
 M151 & --95.39 & 0.79 &                         &           & M238 &  +70.63 & 3.25 & See Sect\,\ref{Sect:ind}&           \\     
 M152 & --109.9 & 0.76 &                         &18 spectra & M240 & --142.4 & 2.91 & See Sect\,\ref{Sect:ind}&           \\          
 M160 & --109.5 & 0.67 &                         & 1 rej     & M250 & --54.72 & 0.59 &                         &           \\            
 M193 & --107.8 & 0.19 &                         &           & M253 & --72.37 & 0.72 &                         &19 spectra \\     
 M199 & --102.8 & 0.47 &                         &           & M256 & --69.14 & 0.73 &                         &19 spectra \\          
 M204 & --110.0 & 0.80 &                         &19 spectra & M264 & --76.72 & 0.58 &                         &           \\         
 M228 & --108.2 & 1.31 &                         &           & M276 & --85.43 & 1.09 &                         &           \\ 
 M230 & --106.2 & 0.79 &                         &           & M285 & --42.35 & 1.13 &                         &           \\ 
 M235 & --103.4 & 0.95 &                         &           & M291 &  +61.45 & 1.05 &                         &           \\                    
 M244 & --105.6 & 3.00 & See Sect\,\ref{Sect:ind}&19 spectra & M295 & --74.93 & 0.57 &                         &           \\ 
 M248 & --111.5 & 0.60 &                         &           & M297 & --56.41 & 0.92 &                         &           \\   
 M252 & --98.51 & 0.76 &                         &18 spectra & M303 &  --47.3 & 0.77 &                         &           \\    
 M255 & --104.8 & 1.11 &                         &           & M304 &   +8.44 & 0.69 &                         &           \\ 
 M258 & --99.11 & 0.63 &                         &           & M320 &  --17.7 & 0.55 &                         &           \\ 
 M260 & --99.67 & 0.75 &                         &           & M321 & --138.0 & 0.76 &                         & 1 rej     \\ 
 M263 & --100.9 & 0.84 &                         &           & M322 &  --49.1 & 2.14 &                         & 18 spectra\\ 
 M267 & --101.8 & 1.04 &                         &           & M323 & --130.6 & 0.66 &                         & 19 spectra\\
 M270 & --108.6 & 0.84 &                         &           & M327 &  --54.9 & 0.66 &                         & 19 spectra\\
 M273 & --105.6 & 0.73 &                         &           & M328 &  --45.3 & 1.46 &                         & 17 spectra\\
 M284 & --106.5 & 1.24 &                         &           & M330 &  --27.1 & 0.59 &                         & 19 spectra\\
 M299 & --105.8 & 1.01 &                         &           & M333 &   +20.3 & 0.66 &                         & 1 rej     \\
 M300 & --106.1 & 1.18 &                         & 19 spectra& M335 &  --65.1 & 0.82 &                         &           \\
 M305 & --97.61 & 0.56 &                         &           & M341 &  --36.5 & 1.96 & See Sect\,\ref{Sect:ind}&           \\
 M308 & --110.3 & 0.71 &                         & 19 spectra& M349 &  --53.0 & 9.11 & See Sect\,\ref{Sect:ind}&           \\
 M319 & --106.6 & 0.84 &                         &           & M356 &  --20.4 & 0.69 &                         &           \\
 M326 & --107.1 & 1.31 &                         &           & M360 & --141.6 & 0.28 &                         &           \\
 M339 & --101.8 & 0.89 &                         & 19 spectra& M362 &   --8.8 & 0.42 &                         &           \\
 M342 & --102.9 & 0.70 &                         &           & M383 &  --61.0 & 0.62 &                         &           \\
 M351 & --106.4 & 0.74 &                         &           &      &                &                         &           \\
 M354 &  --93.3 & 0.92 &                         &           &      &                &                         &           \\
 M357 & --101.5 & 0.87 &                         & 1 rej     &      &                &                         &           \\
\noalign{\smallskip}
\hline
\end{tabular}\newline
\footnotesize{The first and fifth columns list the ID for the star. 
  The second and sixth columns give our final radial velocity for the star and 
  the third and seventh column lists the sigma of the 20 individual measurements
  that the radial velocity is based on. For some stars one  spectrum
  were rejected from the final averaged spectrum. For some stars less than 20 
spectra were available (e.g., due to faulty fibres or miss-placements). The 
number of available spectra is noted in case it less than 20. Stars discussed 
individually in Sect\,\ref{Sect:ind} are also indicated.}
\end{table*}

\begin{figure*}
\centering
\includegraphics[width=8cm]{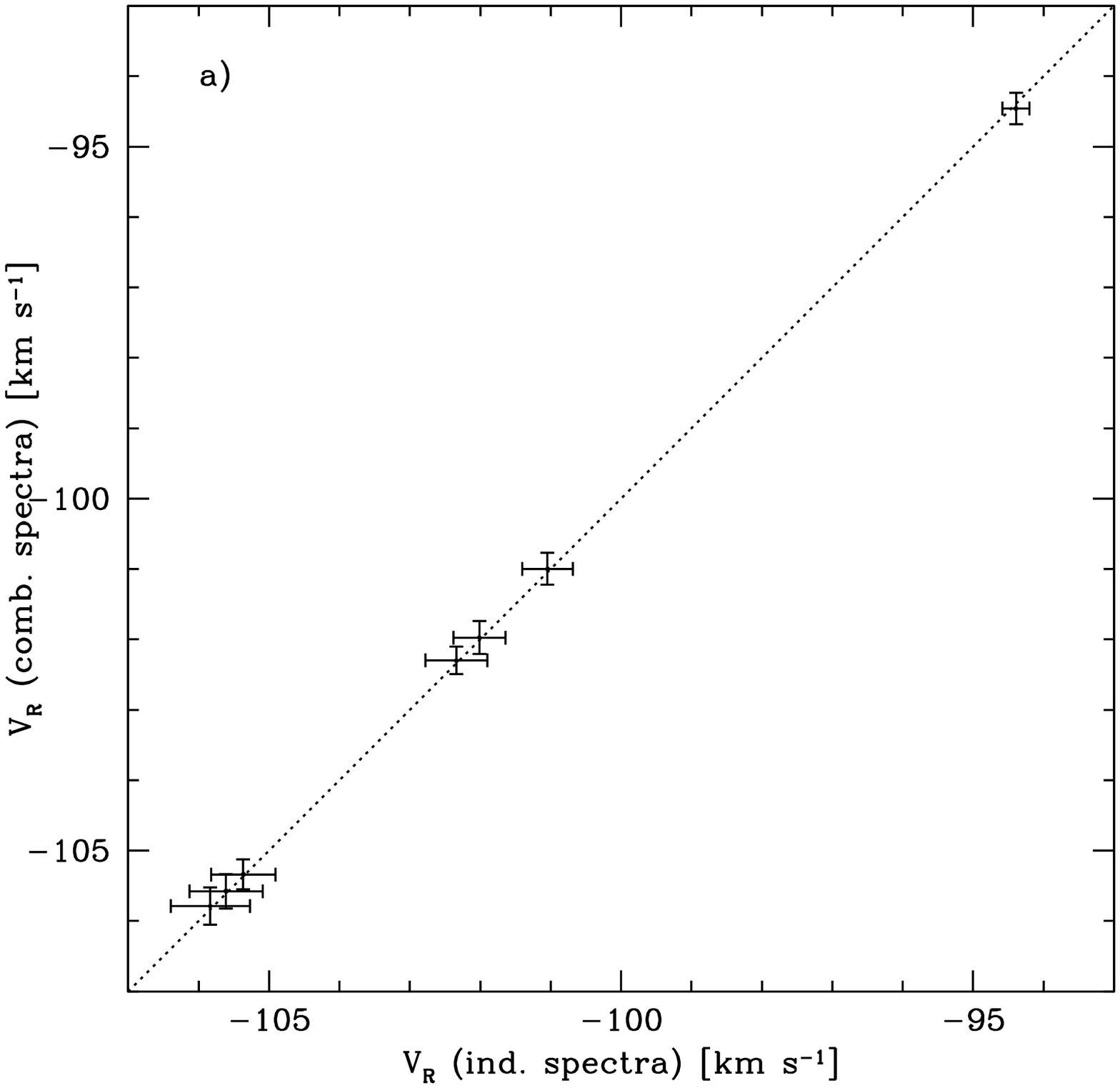}\includegraphics[width=8cm]{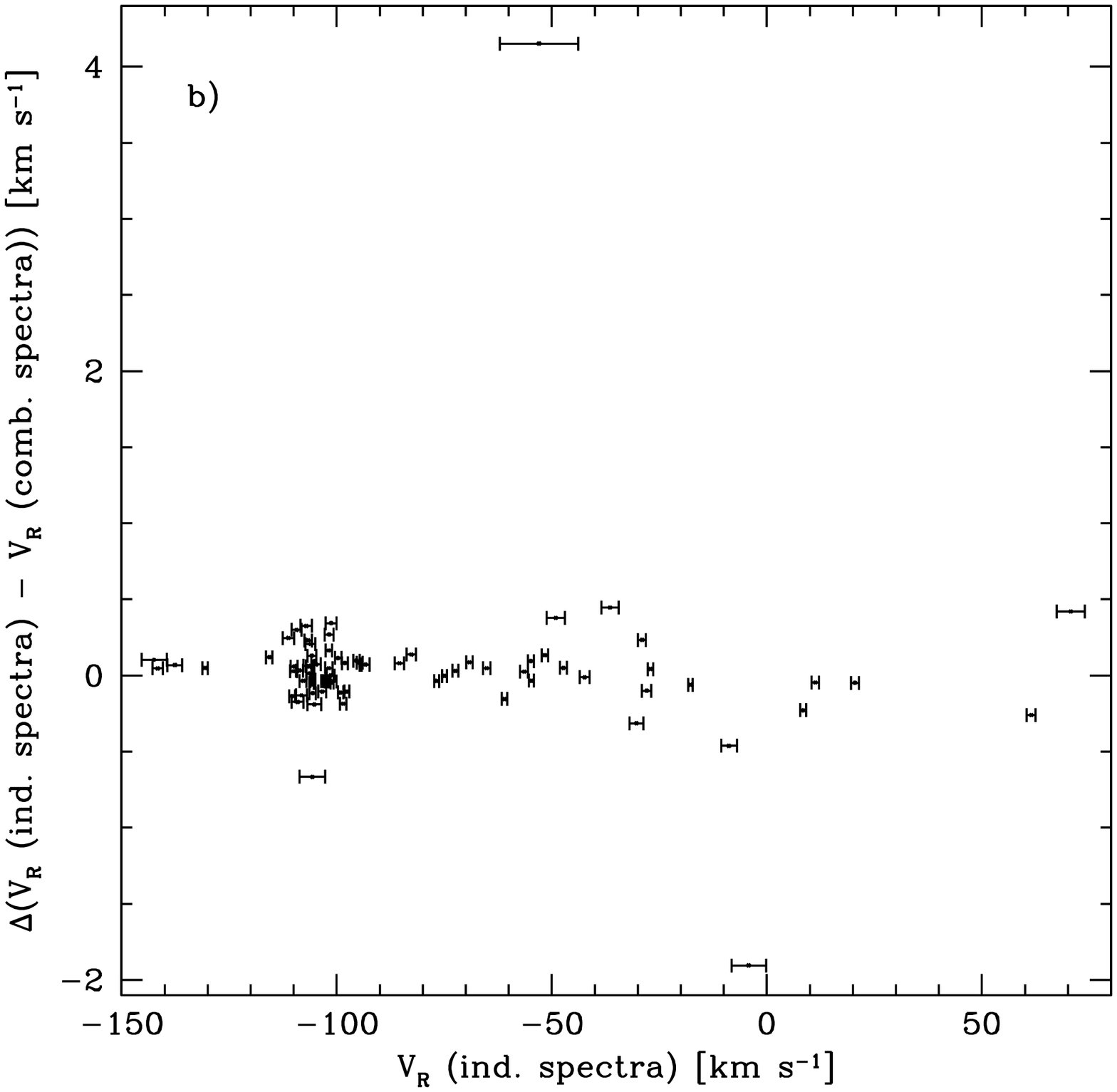}
\caption{{\bf a)} Comparison of radial velocities for the HB/RGB-UVES stars derived from the combined
spectrum ($y$-axis) and from the up to twenty individual spectra separately 
and then averaged ($x$-axis). Errorsbars refer for the $y$-axis to errors
as estimated by {\sc iraf} (see text) and for the $x$-axis to the $\sigma$
around the mean. The dotted line shows the one-to-one relation. {\bf b)}
Shows the difference between the radial velocities for RGB-FLAMES stars derived 
from  the up to twenty individual spectra separately 
and  averaged and the radial velocities derived from the combined
spectra as a function of the radial velocities  derived 
 the  individual spectra and averaged. We only show the 
errorsbars for these values as they are by far the largest. N.B. 
the radial velocities based on the individual spectra are shown
 before the culling leading to the final radial velocities shown
in Table\,\ref{rvrgb.tab}.
}
\label{fig:rverr}
\end{figure*}

Radial velocities were derived using the task {\sc fxcor} in IRAF. The
task cross-correlates a template spectrum with the target spectrum and
reports the velocity difference between the two. Given the wide range
of stellar evolutionary states in the data set, no single template
would adequately serve for all the stars. Both the RGB-UVES and
RGB-GIRAFFE spectra were matched to the observed high-resolution
spectrum of NGC 5927-01 shifted onto a rest wavelength scale.  We
attempted the same procedure for the TO stars, but the very low S/N in
these spectra and the intrinsically greater template
spectrum mismatch made this impractical.  Instead we synthesized a
template spectrum with $T_{\rm{eff}}$=6000 K, log\,{\it g}=4.0, and
[Fe/H]=$-$0.25.

The radial velocities of the TO targets may be expected to be more
uncertain than the other velocities because the synthetic template
contains errors and line omissions due to our imperfect knowledge of
the atomic parameters of all the lines in this region of the optical
spectrum.  Also, the synthetic template does not have exactly the same
wavelength scale.  However, \citet{griffin2000} indicate that radial
velocity errors are greatest when the line broadening in the template
spectrum does not match the target spectrum.  \citet{griffin2000} find
that radial velocity errors introduced by target-template mismatches
are $\pm 0.5$ km\,s$^{-1}$ or less.

The procedure to measure the mean radial velocity for a given star
differs depending on the type of star.  For the HB and RGB candidates
stars we determine the radial velocities in two ways -- first by
cross-correlating a template spectrum with the final, combined
spectrum for the stars and secondly by measuring the radial velocity
for each of the individual 20 OBs, from which we can calculate a mean
and standard deviation. The two methods give extremely similar final
radial velocities but the latter method indicates a larger, and we
would argue more realistic, $\sigma$ (error). Figure \ref{fig:rverr}
shows a comparison between the resulting error-estimates for the two
methods. We note that deriving radial velocities from individual
spectra has the advantage of highlighting stars with variable radial
velocities (see section \ref{Sect:ind}).
We adopt the average over individually measured OBs as our final radial velocity
value.

Table \ref{hb.tab} gives the final radial velocities and error estimates for the
HB/RGB-UVES stars.  Table\,\ref{rvrgb.tab} lists our final radial velocities and
error estimates for the RGB-GIRAFFE stars. 
Some RGB-GIRAFFE stars were not observed across all 20 OBs.  For those
stars observed in every OB, not all stars have 20 spectra for which
{\sc fxcor} would give satisfactory results (e.g., strong cosmic ray
features in a single OB that averaged out in the combined spectrum). A
3$\sigma$ clipping was applied in the final calculation of the radial
velocities for the RGB and HB stars.  For eight RGB-GIRAFFE stars
(M098, M160, M211, M232, M248, M321, M333, and M357) one measurement
was rejected thanks to this clipping scheme (see also
Table\,\ref{rvrgb.tab}). In most cases, once the deviating individual
spectrum was removed the remaining measurements fell within 1$\sigma$.

For the TO stars the individual spectra are too noisy to yield
satisfactory results and we thus only measured their radial velocities
using the final, combined spectra. The results are listed in Table
\ref{to.tab}.  The error reported by IRAF is typically between 1.0 and
1.5 km s$^{-1}$ for the combined TO spectra and between 0.20 and 0.5
km s$^{-1}$ for the RGB-GIRAFFE and HB/RGB-UVES stars. Our tests with
the RGB-GIRAFFE stars indicate that those errors do not reflect the
true uncertainty in the radial velocity measurement over several OBs
for stars observed with the Medusa fibres (see Table \ref{rvrgb.tab}).
However, the IRAF error is consistent with the $\sigma$(OB) for the
UVES stars, where the higher resolving power affords us greater
accuracy (see Table \ref{hb.tab}). An additional source of
  error is the decreasing S/N. Even for the combined TO spectra the
  S/N is low, perhaps resulting in a slightly worse determination of
  the radial velocity. Fortunately, the turn-off velocity
distribution is very clean: the member and non-member stars are very
distinctly separated in radial velocity (see Table \ref{to.tab} and
Fig.\,\ref{fig:rvbins}).  Therefore the precise value of the
uncertainty in radial velocity makes no practical difference for the
TO candidates.

\subsection{Comments on individual stars}
\label{Sect:ind}

\begin{figure}
\centering
\includegraphics[width=8cm]{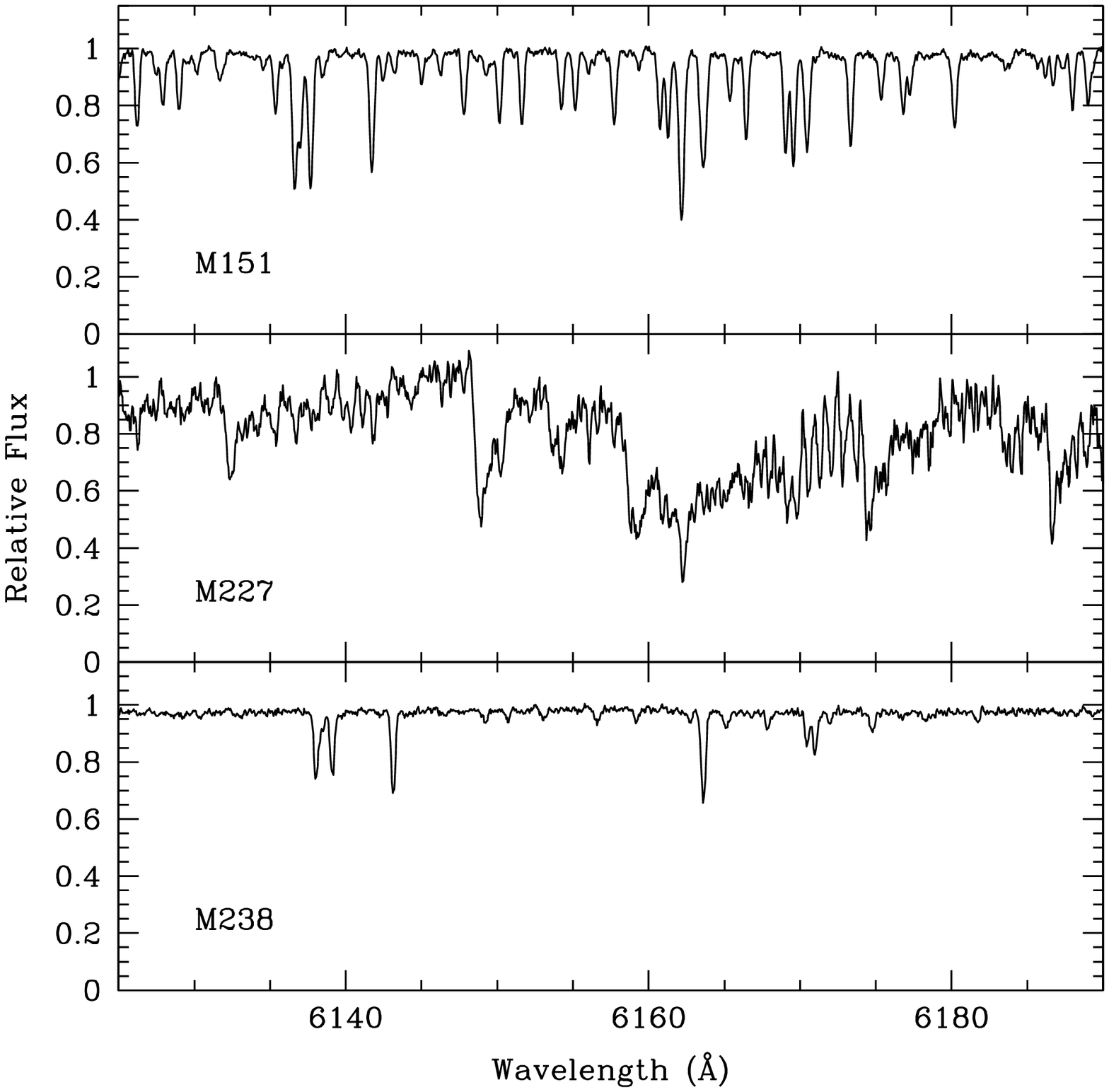}
\caption{Sections of spectra for three RGB-GIRAFFE stars.  The upper
  panel shows M151, which we find to be a radial velocity member of
  NGC\,5927.  The middle panel shows a portion of the spectrum for
  M227, which is a cool foreground dwarf star.  The bottom panel shows
  the spectrum of an RGB star that is significantly more metal-poor
  than the majority of the stars in the cluster. }
\label{fig:ind}
\end{figure}

\begin{figure}
\centering
\includegraphics[width=8cm]{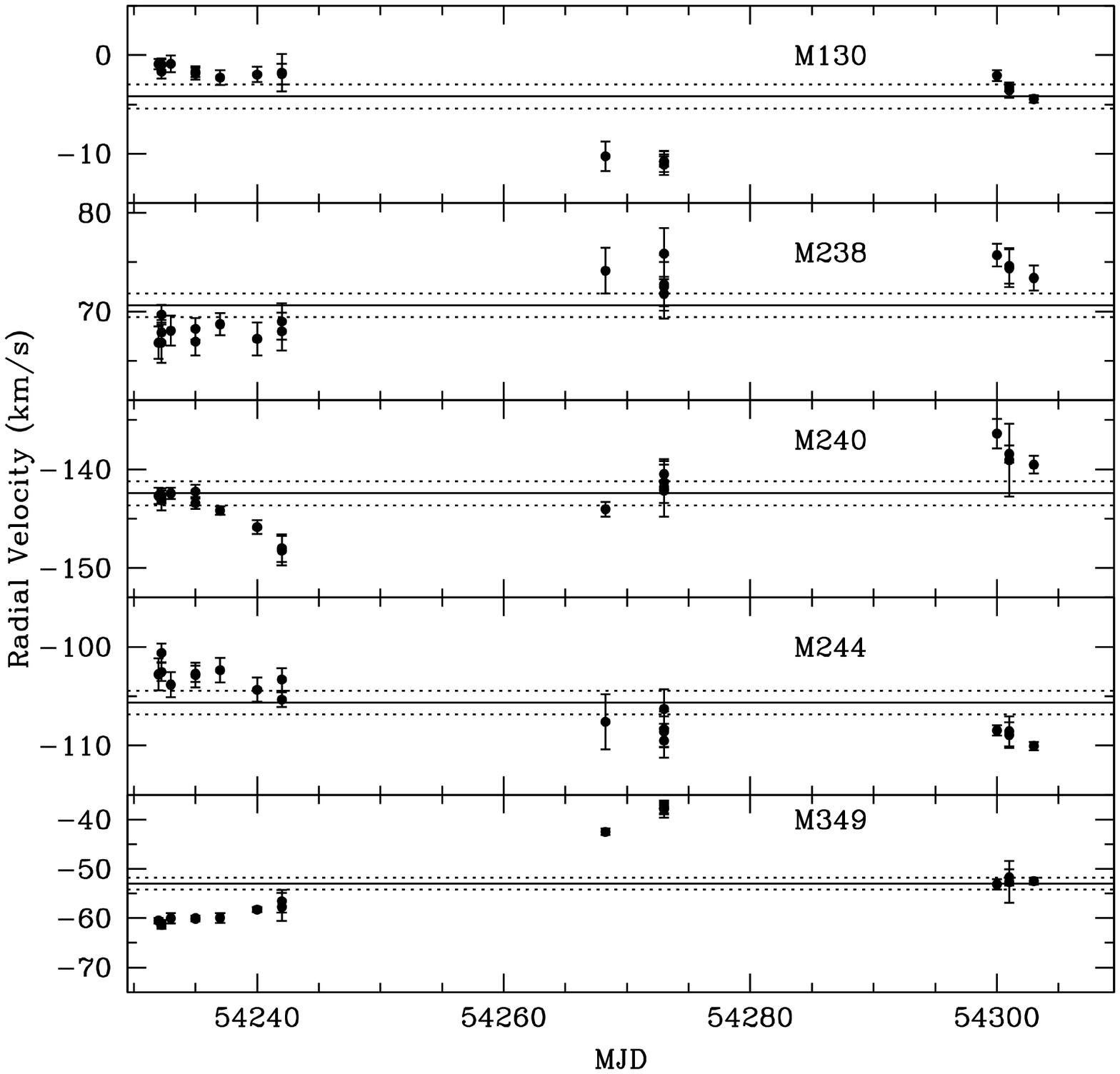}
\caption{The measured radial velocity is shown as a function of the
  modified Julian date (MJD) for five stars that display unusually
  large temporal variations in their measured velocities.  The
  vertical extent of all five panels is $\pm 20$ km s$^{-1}$ from the
  average radial velocity of the star (shown with a solid line). The
  dotted line indicates $\pm$ 1.2 km s$^{-1}$, which is the average
  radial velocity $\sigma$ for the entire RGB-GIRAFFE sample.  Of
  these five stars only M244 has an average radial velocity consistent
  with membership of NGC\,5927.}
\label{fig:altrv}
\end{figure}

\paragraph{M227} This star was selected as a potential RGB
star. Instead of a relatively clean RGB spectrum we have a very rich
spectrum full of molecular band-heads.  Figure\,\ref{fig:ind} shows a
section of the spectrum of M227 and the cluster member star M151 for
comparison.  It was not possible to determine a radial velocity for
M227 with either of our templates.  Given the position of M227 in the
2MASS CMD, it is unlikely that it is an M giant (or that it is behind
the same interstellar extinction as NGC\,5927).  M227 is most probably
an intrinsically red foreground dwarf star.

\paragraph{M238} This star was selected as a potential RGB star. We
note, based on visual inspection of the spectrum, that this star most
likely is quite a bit more metal-poor than the majority of the stars
observed. In Fig.\,\ref{fig:ind} we show a section of the spectrum of
M238 and the cluster member M151 for comparison.  M238 is not a radial
velocity member of NGC\,5927.

\paragraph{M130, M238, M240, M244, and M349} These five stars were
selected as potential RGB stars.  They all show temporal variability
in their radial velocities, as illustrated in Fig.\,\ref{fig:altrv}.
Our observations are clumped in time.  For M130, M238, and M244 there
is not much variation within each group of observations carried out
close in time, but the variability shows up on longer timescales. M349
and M240 have the potential for variability on a much shorter time
scale.  The difference between maximum and minimum velocity is around
10 km\,s$^{-1}$ for all stars but M349, where the difference is as
large as 20 km\,s$^{-1}$.  All five stars show a radial velocity
variation at least 2.5 times larger than the average standard
deviation across all 20 OBs for all our RGB-GIRAFFE targets
($\sigma=1.2$km\,s$^{-1}$).

Since the fibre configuration was not constant across all 20 OBs
we have investigated the possibility that stars were misidentified,
either through improper fibre positioning or faulty fibre-star
matching.  In all cases fibre coordinates were well matched to the
correct stellar coordinates.  Furthermore, fibre configuration changes
do not correlate with the observed radial velocity variations.  Radial
velocity variations as large as these typically indicate stellar
companions. There is no indication of a secondary spectrum in any
of these cases.  We note that of these five potential binary systems
only M244 is a radial velocity member of NGC\,5927 (see
Table\,\ref{rvrgb.tab}).

\paragraph{M341 and M322} These stars were selected as potential RGB
stars.  They have a very large $\sigma$ for the radial velocities when
all OBs are considered, but radial velocity measurements from
individual OBs fall within 3$\sigma$ of the final value. We have not
been able to identify anything that might explain the large
$\sigma$. Neither star is a radial velocity member.

%
\section{Cluster membership and the radial velocity of NGC\,5927 \label{memsec}}

\begin{figure}
\centering
\includegraphics[width=8cm]{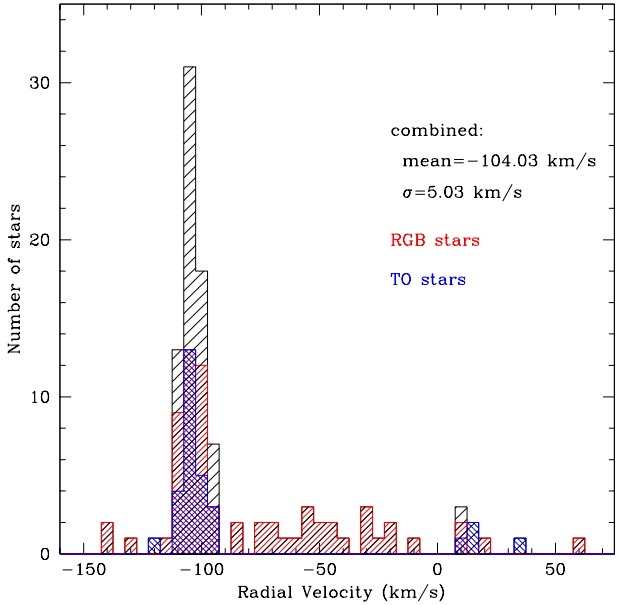}
\caption{The distribution of measured heliocentric radial velocities
  for our target stars.  Radial velocity variables are excluded.  The
  black histogram is the combined sample.  The red and blue histograms
  are the RGB-GIRAFFE and TO samples, respectively.}
\label{fig:rvbins}
\end{figure}

\begin{table*}
\caption{Radial velocities for NGC\,5927 from the literature.
}
\label{TabHBrvref}
\setlength{\tabcolsep}{0.03in}
\renewcommand{\footnoterule}{}  
\begin{tabular}{rrrlllll}
\hline \hline
\noalign{\smallskip}
\multicolumn{1}{c}{Radial velocity}     & \multicolumn{1}{c}{$\sigma$}        & No. of objects & Type of obs.  &  Reference &   \\
\multicolumn{1}{c}{[km s$^{-1}$]}      &        \multicolumn{1}{c}{[km s$^{-1}$]}                       \\
     \noalign{\smallskip}
     \hline
     \noalign{\smallskip}
--96 &$\pm$ 28  &  & Integrated light & \citet{1959MNRAS.119..157K}                   \\
--78 & $\epsilon$=12 & 7 &  Compilation      & \citet{1981ApJS...45..259W}    & $ \dagger$              \\
--74 &$\pm$  15  &   6  & Individual stars              & \citet{cohen1983}     &         \\
--93 & $\pm$ 14  &  & Integrated light & \citet{zw84}           \\
--106 &  $\pm$ 7 & &Integrated light & \citet{1986PASP...98..403H}\\
--113 & &  & Integrated light & \citet{1988AJ.....96...92A} (no error)   \\
--106.6& $\pm$ 5.2&   8 & Individual stars                 & \citet{suntzeff1993}  & $\dagger \dagger$       \\
--118.6& $\pm$ 8.8&   5 & Individual stars                 & \citet{1995AJ....109.2533D}  \\
--123.8& $\pm$ 8.8&   19 & Individual stars                & \citet{1997PASP..109..883R}  \\
--84& $\pm$ 5  &   20 & Individual stars                & \citet{carrera2007}   \\
--110.3 &$\pm$ 21.5  & 43 & Individual stars  & \citet{2010AA...524A..44P} \\
--104.03&$\pm$ 5.03  &  70& Individual stars                 &   This work             \\
      \hline
      \end{tabular}\newline
\footnotesize{The first column lists the heliocentric radial velocity. The second
column lists the number of stars used in the derivation of the mean
velocity of the cluster. For four studies integrated light has been
used instead, as indicated. Column three lists the reference for 
the measurement. Any additional comments are given in column four.\newline
$\dagger $ For a description of the calculation of the total error, including
stellar variability, see \citet{1981ApJS...45..259W}.\newline
$\dagger \dagger$ Average excludes star M468 and star A.
}
\end{table*}

Figure \ref{fig:rvbins} shows our derived velocity histograms for the
heliocentric radial velocities of the targeted stars. Stars showing
radial velocity variability have been excluded (see section
\ref{Sect:ind}).  It is clear that the RGB-GIRAFFE sample in the outer
part of the cluster contributes most of the background/foreground
contamination. The majority of the TO candidates are members of the
cluster.

The relatively large number of RGB-GIRAFFE candidates and the
relatively high fraction of non-members make an iterative rejection
necessary.  First, we considered only stars with --150 km\,s$^{-1} <
$v$_r < -70$ km\,s$^{-1}$, since it is clear that the systemic
velocity for the cluster must be in this range.  Next, we calculated
the mean and standard deviation ($\sigma$) of the radial velocities
for all target stars with measured radial velocities in that range.
Then, we excluded stars more than 3$\sigma$ from the initial mean
value.  Finally, we re-calculated the mean velocity from the culled
sample. We iterated in this way until we could exclude no more stars
with a 3$\sigma$-clipping. We need 4 further iterations to reach the
final value. In this way we derive a mean velocity for NGC\,5927 of
--104.03 km\,s$^{-1}$ with $\sigma$ =5.03 km\,s$^{-1}$.

This resulted in the 72 stars (RGB-UVES, RGB-GIRAFFE, and TO) that
satisfy the membership criteria.  All of our RGB-UVES targets are
radial velocity members and 26 of our 30 TO candidates are radial
velocity members.  Our RGB-GIRAFFE candidate success rate is much
lower: only 39 of our 75 target stars are 3$\sigma$ members of
NGC\,5927, i.e. a 50\%.

Since NGC\,5927 lies in the plane of the Milky Way foreground
contamination in the cluster outskirts is quite likely.  We have much
greater success in the inner 5''.  Radial velocity studies with small
numbers of stars or limited to the outer regions of the cluster thus
may suffer from a large number of non-member stars.  If we had been
limited to target selection from ground-based photometry (i.e. not
accessing the crowded core) and/or had had fewer targets overall, it
is quite likely our systemic radial velocity would have skewed to
lower values, where the bulk of the foreground contamination lies.

There are several determinations for the radial velocity for NGC\,5927
in the literature (see Table\,\ref{TabHBrvref}). One third of the
available estimates are based on integrated light rather than single
stars.

The spread in radial velocities within a single study can be quite
large. For example, although \citet{1959MNRAS.119..157K} reports an
average radial velocity of --96 km\,s$^{-1}$ (based on 6 stars)
individual measurements range from --204 km\,s$^{-1}$ to +3
km\,s$^{-1}$. Two stars fall at --118 km\,s$^{-1}$ and two close to
--70 km\,s$^{-1}$.  We suspect that their sample is heavily
contaminated with foreground and background stars.
\citet{1981ApJS...45..259W} is a compilation based primarily on
\citet{1959MNRAS.119..157K} but with the addition of a single Mira
variable from \citet{1974Obs....94..133A}. \citet{1981ApJS...45..259W}
reports a weighted average of radial velocity values but contamination
can not be excluded.

\citet{cohen1983} reports radial velocities from individual stars.
Her mean velocity is low compared to almost all other
determinations.  However, \citet{cohen1983} does report instrument
difficulties that limited the internal measurement accuracy to 15
km\,s$^{-1}$.  The scatter around the mean value is large,
$\sigma=$15 km\,s$^{-1}$.

We note that the value from \citet{carrera2007} differs the most (when
considering results based on individual stars). A likely explanation is
that these observations were taken with FORS2 on the VLT, i.e., a low
resolution ($R\sim 5\,000$), multi-slit spectrograph where
misalignments between stars and slits hampered the determination of
the radial velocities. This offset can produce differences in radial
velocities of about 15\,km\,s$^{-1}$ (the $\sigma$ quoted in
Table\,\ref{TabHBrvref} is the scatter around the mean value). A 
realistic error in the radial velocity would be 10 -- 15\,km\,s$^{-1}$
(Carrera, {\sl priv. com.}).  \citet{carrera2007} were primarily concerned
with measuring the strength of the Ca\,{\sc ii} triplet lines and
hence this was not problematic for their science case.

The difference can likely be traced to that the observations in
\citet{carrera2007} were obtained in MXU mode (multi-slit) and not
long-slit. Since the masks are created prior to the actual
observations it can be difficult to ensure that all stars are properly
placed on the slits resulting in decreased accuracy for the radial
velocities and can account for the observed discrepancies (Carrera,
{\sl priv. com.}). This can become a significant problem, e.g., if the
seeing is smaller than the actual slit width. That this is a potential
problem for observations with the FORS instruments has been previously
reported in the literature \citep[e.g.,][]{irwin2002}. We note that
for \object{NGC\,6532} \citet{feltzing2009} derive a radial velocity
$\sim$\,20\,k\,ms$^{-1}$ larger than the value from
\citet{carrera2007}. Again, likely resulting from the same cause.

\citet{2010AA...524A..44P} obtained velocities for a large number of
stars in NGC\,5927. Our value agrees well with theirs within
the error-bars.  Finally, we note that of all the single-star radial
velocity measurements in the literature, only \citet{suntzeff1993}
reports having observed stars that are not radial velocity members and
we have excluded those from the average in Table\,\ref{TabHBrvref}.
Our average radial velocity agrees very well their value.

%
\section{Summary}

We have obtained spectra of 105 stars in the direction of the 
metal-rich globular cluster NGC\,5927 using the UVES and GIRAFFE
fibres in the FLAMES spectrograph. Based on the measured radial
velocities we  identified the following members of NGC\,5927:

\begin{itemize}

\item 7 HB and RGB stars with high-resolution spectra 

\item 39 HB and RGB stars with medium-resolution spectra

\item 26 TO stars with medium-resolution spectra

\end{itemize}

\noindent
We also found:

\begin{itemize}
\item one foreground cool dwarf star,
\item  one likely very metal-poor,
giant or subgiant star,
\item five radial velocity variables with no
companion star visible in the spectrum of the primary. One of these variables
 is a radial velocity member of NGC\,5927.
\end{itemize}

Our final radial velocity, v$_r$=--104.03$\pm$5.03 km s$^{-1}$, agrees well
with the majority of most recently published values, but is based on a
much larger sample of high-quality spectra.


\begin{acknowledgements}
We thank the referee, R. Carrera, for a careful read of the manuscript
and valuable suggestions that increased the readability of the paper. We would
in particular like to thank him for the useful information on the FORS2 
observations (Carrera et al. 2007) which gives a likely explanation 
to the discrepancy between our and their values for the radial velocities.

We would like to thank Rachel A. Johnson for her work in the
identification of target stars for this observing program. We would
also like to thank Thomas Brown for providing the ACS data.  JS
acknowledges the support of an Incoming International Marie Curie
Fellowship. SF is a Royal Swedish Academy of Sciences Research Fellows
supported by a grant from the Knut and Alice Wallenberg Foundation. SF
acknowledges support from grant number 2008-4095 from the Swedish
Research Council. This work has made use of the NED, SIMBAD, and VALD
databases. This publication makes use of data products from the Two
Micron All Sky Survey, which is a joint project of the University of
Massachusetts and the Infrared Processing and Analysis
Center/California Institute of Technology, funded by the National
Aeronautics and Space Administration and the National Science
Foundation.
\end{acknowledgements}

\bibliographystyle{aa}
\bibliography{referenser}

\end{document}